**Title:**

Comparison of Spatiotemporal Characteristics of Eye Movements in Non-experts and the Skill Transfer Effects of Gaze Guidance and Annotation Guidance

**Short title:**

Exploring Eye Movement Characteristics and Skill Transfer: Gaze vs. Annotation Guidance for Non-experts

**Authors:**

Shota Nishijima[1,2], Asuka Takai[1,2]

**Affiliations:**

[1] Advanced Telecommunications Research Institute International (ATR)

[2] Osaka Metropolitan University

**Abstract:**

　　Methods for converting experts' tacit knowledge—such as areas of focus and judgment criteria—into explicit knowledge that can be conveyed through text or diagrams have drawn increasing attention. Gaze data has emerged as a valuable approach in this effort. However, the effective transfer of such tacit knowledge remains a challenge. No studies have directly compared the effects of gaze-based and annotation-based guidance or adequately examined their impacts on skill improvement after instruction. In this study, we examined the effects of gaze and annotation guidance on the spatiotemporal characteristics of eye movements (fixation duration, fixation count, and fixation area) and the skill transfer of expert evaluation techniques in *karate kata* performances.

　　We recruited 28 non-expert participants and assigned them to three groups: a gaze guidance group, an annotation guidance group, and a control group. Participants were presented with instructional slideshows based on the expert's gaze data and annotations. Before and after instruction, participants were asked to evaluate *karate kata* performance videos performed by practitioners with different skill levels (*dan* ranks). Changes in ranking results were analyzed. A machine learning model was employed to automate the segmentation of observation areas and convert gaze coordinate data into the coordinate system corresponding to the video to evaluate eye movements. The automatic process greatly streamlined the analysis process. Fixation duration, fixation count, and the distribution of fixations by body areas were analyzed for three characteristic movements in the videos.

The results showed that the annotation guidance group exhibited an effect of directing gaze toward the presented areas of focus, with a trend of increased total number of fixation areas (Cohen's d = 0.2 to 0.4). On the other hand, while the gaze guidance group was not encouraged to make fixations on multiple areas, the possibility of promoting peripheral vision was inferred based on measurements from the eye-tracking system. Regarding ranking results, 71.4% of participants in the gaze guidance group showed an improvement in ranking skills, with a trend toward better scoring ability compared to the other groups.

This study provides the first comparison of instructional methods and their effects on skill improvement, demonstrating the necessity of selecting appropriate methods based on instructional goals to transfer tacit knowledge effectively. Gaze-based instructional methods are expected to be versatile and applicable to not only *karate kata* but also fields such as industry, medicine, and other sports. Further validation of Gaze-based instructional methods may enable a more efficient transfer of expert skills across diverse fields.

# 1. Introduction

## 1.1 Background

In recent years, efforts to understand and apply expert skills have been explored across various fields, such as industry, healthcare, and sports. For example, in the industrial field, digital manufacturing has been proposed, which involves transforming tacit knowledge into explicit knowledge and further converting it into digital values to maximize the use of information technologies in production processes[1]. Explicit knowledge refers to knowledge that can be articulated through text or diagrams. In contrast, tacit knowledge is knowledge that cannot be easily verbalized and requires significant effort and experience to acquire, often considered the most subjective and transparent form of knowledge[1-5]. In the healthcare field, attempts have been made to convert tacit knowledge into explicit knowledge through dialogue and description in individual cases, such as one-on-one interviews with radiological technicians[6] or participatory observations and interviews with caregiving staff during practical tasks in dementia services[7]. In the sports field, efforts to quantify motion data using sensors to extract characteristic features and identify specific skills[8], quantifying subjective evaluations from coaches[9], and comparing the distinguishing features observed in experts and novices[10] have been attempted.

The industrial field has advanced automation using Artificial Intelligence because it allows direct approaches to the state of physical objects. For example, welding tasks requiring advanced manual skills have been targeted for automation through image recognition and robotic control technologies[11]. In contrast, the healthcare and sports fields require not only processing tacit knowledge as data but also effectively transferring this knowledge to non-experts. Therefore, there is a growing demand for improving instructional methods. Explicit knowledge is relatively easy to transfer and has traditionally been conveyed through text or diagrams. In modern times, this has been further facilitated by manuals and mechanization. However, the transfer of tacit knowledge still primarily relies on on-the-job training, where direct instruction occurs in practical settings or through long-term experiential learning by the learners themselves. Off-the-job training using video-based instruction has also been proposed as an alternative method for efficiently transferring tacit knowledge to a broader audience.

## 1.2 Related Studies

In recent years, numerous studies have been conducted using eye-tracking methods to extract and visualize the tacit knowledge, such as intuition and know-how, of skilled professionals[12-16].

For example, Morita et al. investigated quality control in construction sites by combining eye-tracking measurements and interviews[12]. They aimed to clarify the viewpoints, situation recognition, and judgment criteria that lead to actions by skilled field supervisors. As a result, they discovered key viewpoints through eye-tracking measurements, and by combining eye-tracking and interviews, they identified the key viewpoints of skilled supervisors and reported the possibility of interpreting viewpoints via interviews instantly. Furthermore, their method suggested that the extracted behavioral process could be utilized for skill transfer to less-experienced supervisors. However, this study did not test the effectiveness of applying eye-tracking to actual teaching or educational programs but merely indicated its potential for skill transfer.

On the other hand, Sasamoto et al. applied eye-tracking in an educational program[17]. They developed and tested a short educational program for teaching transfer movements in caregiving. The program combined instructional videos that displayed eye movements with annotations and practical training. The study focused on gaze, an eye movement that acquires external information. The results showed that using the program led to a decrease in instantaneous gaze and an increase in deliberate gaze, with an overall reduction in total gaze count. This finding suggested a shift from a superficial understanding, e.g., attention-level understanding, to a more conscious, thought-driven understanding.

Fukui et al. analyzed gaze tendencies when ski instructors and beginners identified specific points requiring correction[16]. They found that beginners exhibited longer gaze movements, often shifting between their feet and heads, which indicated scattered attention. In contrast, instructors with experience, who had received multiple lectures from professional instructors, focused only on the points pointed out by the professionals. However, they failed to identify necessary corrective points. This result suggested that identifying corrective points requires looking at multiple areas intentionally rather than focusing on a single point.

From the above studies, while eye-tracking has shown potential for enhancing education, it has been suggested that deliberate efforts to direct learners to multiple key viewpoints are necessary. While Morita et al. demonstrated the potential of eye-tracking to extract the viewpoints of experts[12], the actual effectiveness of applying these findings to educational programs has not been verified. On the other hand, Sasamoto et al. proposed an educational program using eye-tracking[17]; however, this program combined multiple instructional methods, such as displaying eye movements, providing annotations, and actual training. As a result, the individual effects of each instructional method remain unclear when used independently. This highlights a gap in the current research, where the specific effectiveness of distinct instruction methods, such as eye-movement-based instruction and annotation-based instruction, has not yet been fully examined. Moreover, the effectiveness of using eye-tracking to indicate key viewpoints for skill transfer has not yet been fully verified.

## 1.3 Research Objectives

The ultimate goal of this study is to develop a method for conveying skills that can be applied universally across various fields, thereby facilitating the transfer of expert skills. To achieve this, it is necessary to examine methods for effectively conveying tacit knowledge, specifically the key viewpoints experts focus on, which have not been fully addressed in previous studies. Specifically, the study aims to clarify how different instruction methods influence the spatiotemporal characteristics of eye movements and to evaluate how closely non-experts can approximate the observation and judgment of experts.

In this study, the expert's tacit knowledge will be extracted based on previous research[12], and two teaching methods—instruction using eye movement and instruction using annotations—will be compared to verify their effectiveness. Furthermore, the study focuses on scoring-based sports, where subjective evaluation can result in variability, rather than sports in which objective measures, such as scoring in ball games, determine the outcome. In particular, this study targets *karate kata* competition, where no physical contact with an opponent and no external factors, such as equipment like balls, affect the performance. Judging the rapid sequence of body movements requires instantaneous evaluation and expert skills. Recently, *karate* experts have increasingly shared key points for scoring, promoting converting tacit knowledge into explicit knowledge[18, 19]. This makes *karate kata* an appropriate subject for this study. Additionally, in 2018, the judging system was changed from the traditional flag-raising method to a scoring system with predetermined criteria, ensuring that the evaluation items are clearly defined even though it is a sport based on visual evaluation.

Moreover, *karate* has gained significant attention due to the mandatory inclusion of martial arts and dance in junior high school physical education in Japan since 2012 and its adoption as an Olympic sport in 2020. However, from the perspective of skill transfer, a lack of instructors remains challenging in educational settings. According to a survey[20], reasons for not incorporating *karate* into physical education classes include a lack of confidence in teaching skills and difficulty securing instructors. This highlights the need for accessible and high-quality training programs and teaching materials for teachers. In other words, there is a demand for instruction methods that contribute to developing referees and instructors.

Based on the above, this study focuses on *karate kata* competition and uses two distinct instruction methods—instruction using eye movement and instruction using annotations—to guide non-experts in observing various aspects of the performance. Changes in the spatio-temporal characteristics of eye movements before and after instruction will be investigated. Additionally, a ranking task, in which participants assign correct rankings to *karate kata* performances by performers of different ranks, will be used to evaluate changes in scoring ability. Through this process, the study will assess the effectiveness of the instruction methods for skill transfer.

## 2. Methods

The experiment was conducted with 28 male and female participants with little or no prior experience in *karate* (including beginners and inexperienced individuals). One experienced practitioner, a *Shihan* (coach) from the *Nippon Karate-do Taishikan* (Japan Karatedo Federation *Goju-kai* 5th *dan* and official referee), also participated in the study. Written informed consent was obtained from all participants prior to their participation, and the research protocol was approved by the Ethics Committee of the Advanced Telecommunications Research Institute International (ATR).

## 2.1 Experimental Methods

### 2.1.1 Experimental Setup

The experimental setup and its schematic diagrams are shown in Fig. 1 and Fig. 2. The experimental setup consists of an eye-tracking device (EMR: Eye Mark Recorder), a chinrest, a monitor (27-inch), and personal computers for data recording and video playing. Participants wore the EMR, sat on a chair, and observed the images displayed on the monitor while their head movement was fixed using the chinrest. The distance between the chinrest and the monitor is 1 meter. A black curtain is placed behind the monitor to ensure that the participant's field of view does not include any distractions, such as the appearance of the room behind the monitor. This curtain is large enough to block any unnecessary visual elements and is intended to eliminate factors that might disrupt the participant's concentration during video observation. A black cloth is also draped over the table to prevent reflections of light or other unwanted visual stimuli from entering the participant's view from the table's surface. For the experiment involving an expert, as described later in Section 2.1.5, verbal comments must be recorded. Therefore, to capture video and audio recordings, a webcam was installed behind the participant's seat, positioned so that both the expert and the monitor were within the same frame.

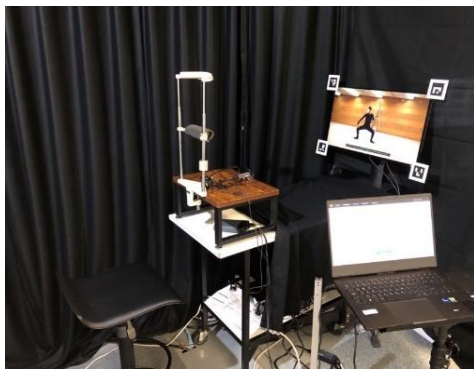 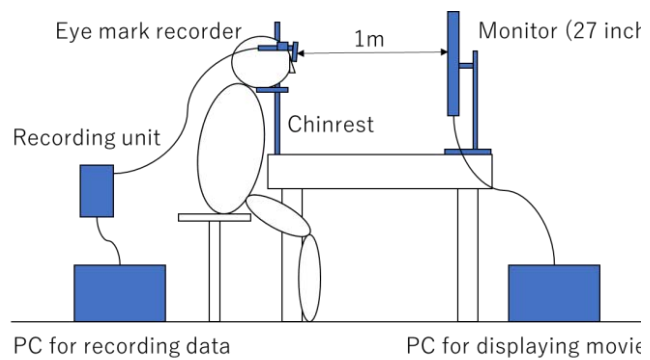

Figure 1 Experimental setup       Figure 2 Schematic of experimental setup

This experiment used wearable EMRs, namely Tobii Pro Glasses 2 and Glasses 3, manufactured by Tobii Technology Co., Ltd.. If participants required vision correction with glasses or contact lenses, prescription lenses provided as accessories with the EMR were used to ensure the video was seen clearly. The clarity of vision was determined based on the participants' subjective judgment.

The data obtained by the EMR is initially stored in a recording unit and then exported to a personal computer for data recording using the software provided with the EMR (Tobii Pro Glasses Controller). The sampling rate was 100 Hz. The gaze data was output using Tobii Pro Lab, a software provided by Tobii Technology Co., Ltd., and analyzed using a custom-made program described later in Section 2.2.

### 2.1.2 Experimental procedure (non-experts)

The outline of the procedure for the observational experiment with non-experts is shown in Fig. 3. The experiment begins with observing a sample performance video. Since the participants are non-experts with no or little prior knowledge of *karate kata*, they may not understand which movements to use as criteria for ranking when viewing them for the first time. Therefore, before observing the performance videos subject to ranking, participants are shown a sample performance video recorded in the same environment. This process is supposed to help participants understand the types of movements involved in *karate kata*, eliminate the influence of novelty arising from seeing the movements for the first time, and serve as a reference for subsequent observation and evaluation.

Next, pre-recorded performance videos are observed while participants' gaze are tracked using the EMR. The details of the performance videos used in this stage are described in the next section. After the observation, participants were asked to write on a questionnaire whether they felt the performance was skillful. Observing the performance videos and filling out the questionnaire constitute one set, and a total of four sets are conducted for four performance videos. During the observation, the weight of the EMR may cause slight displacement on the face, and the position of the device may also shift during questionnaire completion or when using the chinrest. Therefore, the EMR is recalibrated at the beginning of each set to maintain high measurement accuracy. After watching each video, participants rate it on a 5-point subjective scale and provide reasons for their ratings in a free-text section on the questionnaire.

Subsequently, participants determine the rankings of the four performance videos. At this stage, ties are not allowed, and participants are instructed to assign each video a unique rank from first to fourth place. Participants can refer to their evaluation scores and comments written on the questionnaire and use the sample performance video as a benchmark for determining the rankings.

Subsequently, participants were given instructions regarding the key points highlighted by an expert. After they had observed the four videos again, they completed the questionnaire and determined the rankings.

Additionally, after the experiment, a verbal interview was conducted to determine the points of focus during the first and second rounds of observation to confirm whether the instructions were applied during the second round.

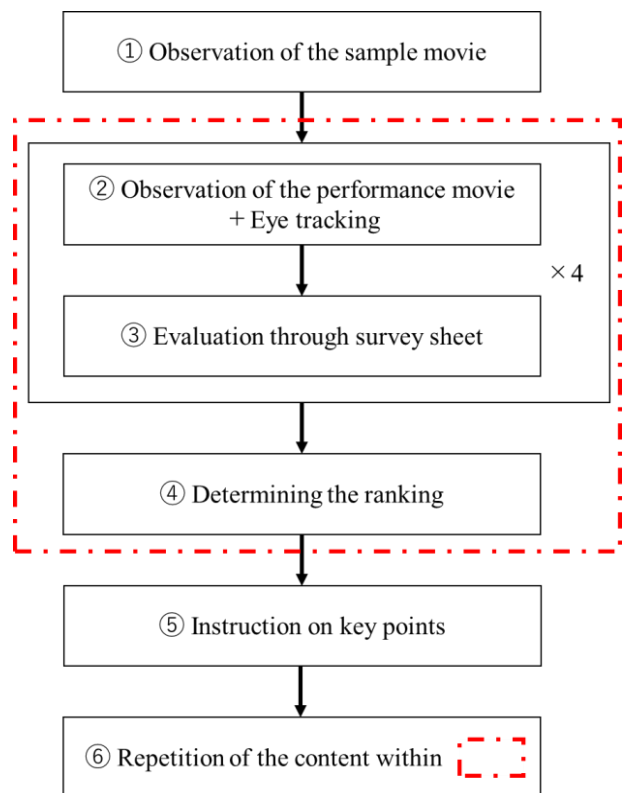

Figure 3 Protocol of experiment for non-experts

The non-experts were divided into three groups. Details of the instructions based on the expert's gaze data and evaluative comments are described in Section 2.1.4. For the first group, the expert's gaze points were plotted on snapshots of the performance videos and shown to the participants; this group was referred to as the gaze group. For the second group, brief explanations or keywords were extracted from the expert's evaluative comments and annotated on snapshots of the performance videos; this group was referred to as the annotation group. However, to ensure parity in the amount of information provided for ranking purposes between the gaze group and the annotation group, the explanations and keywords displayed for the annotation group excluded any evaluative comments or indications of skill level, focusing solely on areas of interest. The third group was given no instructions and was referred to as the control group.

During the experiment, there were occasional difficulties in maintaining data acquisition rates for eye movement measurement, making it challenging to continue the experiment. Although the exact cause was unclear, it was presumed to be eye fatigue or loss of concentration from prolonged observation. To address this, a 15-minute break was provided after the first round of ranking. Additionally, when data acquisition rates declined, a 10-minute break was introduced between sets as needed, regardless of whether participants reported fatigue or loss of concentration.

### 2.1.3 Performance Videos

Fig. 4 shows scenes from one of the performance videos presented during the observation experiment. The selected *kata* was *Gekisai Dai Ichi*, taught as a fundamental *kata* in the *Goju-ryu* style at the Japan *Karate-do Taishikan*. This *kata* was chosen because it is a form commonly learned and performed by *karate* practitioners regardless of their skill level. Five *karate* practitioners with different ranks were asked to cooperate as performers. The ranks of the performers were as follows: one 8th *kyu*, three 1st *dan*, and one 2nd *dan*. In *karate*, the ranking begins at 10th *kyu*, and as proficiency increases, the *kyu* number decreases. After 1st *kyu*, the ranks advance to 1st *dan*, 2nd *dan*, and so on.

The duration of the performance varied slightly among the practitioners but was approximately one minute for all. Among the five performers, the video of one 1st *dan* practitioner was used as the sample performance video shown to the participants at the beginning of the experiment. The observation experiment used the performance videos of the other four practitioners.

Traditionally, *karate* performances are conducted while wearing *dogi* (white uniforms) and *obi* (belts) tied around the waist. Observers can discern subtle body movements not only through the movements of the performer's body but also through changes in the shadows on the *dogi* and the movement of the *obi*. However, for this experiment, all performers wore black bodysuits to enhance the accuracy of body part estimation (explained in Section 2.2.1) by clearly defining the shape of each body part. Additionally, this choice aimed to prevent participants from inferring the performers' skill levels based on their attire.

Moreover, breathing, which is an element typically considered in scoring, was excluded from evaluation in this study. This exclusion was intended to avoid participants estimating the performers' rankings based on their perceived age, as age might indirectly indicate proficiency. To further ensure anonymity, the performers' faces were pixelated to prevent recognition of their age or identity.

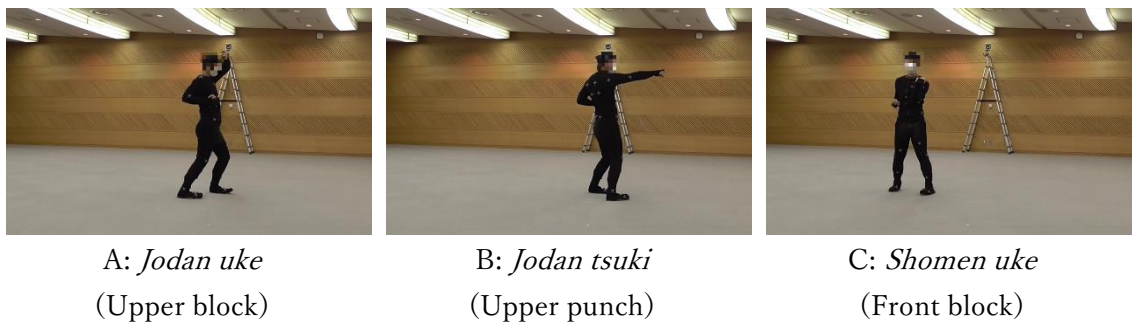

| A: *Jodan uke* | B: *Jodan tsuki* | C: *Shomen uke* |
| (Upper block) | (Upper punch) | (Front block) |

Figure 4 Evaluated scenes from performance movie

### 2.1.4 Instructional Slideshow

Based on an experiment involving an expert *karate* practitioner conducted using the same performance videos (see Section 2.1.5), the results were used to create instructional slideshows for the gaze and annotation groups. The reason for using slideshows instead of videos for instruction is based on the findings of a web usability study[21]. The study compared Retrospective Think Aloud (RTA) and found that gaze plot-cued RTA (eye movements superimposed on still images) was nearly as effective as gaze video-cued RTA (eye movements superimposed on screen videos), excelling in eliciting visual and cognitive comments while identifying the same number of usability problems.

Thus, the present study determined that slideshows featuring still images with gaze plots are more suitable for instruction than videos. Note that as the instructional slides are based on still images, this study does not analyze temporal aspects of gaze data.

The slideshow consists of 21 still images representing individual actions within the *kata*, such as *Tsuki* (punch), *Uke* (block), and *Keri* (kick). Extremely short actions were combined with adjacent actions to form single units. For the gaze group, the slideshow was created by overlaying red circles on the still images to indicate gaze points from the expert practitioner's gaze data (Fig. 5). Each image was displayed for three seconds, approximating the playback duration of the original video.

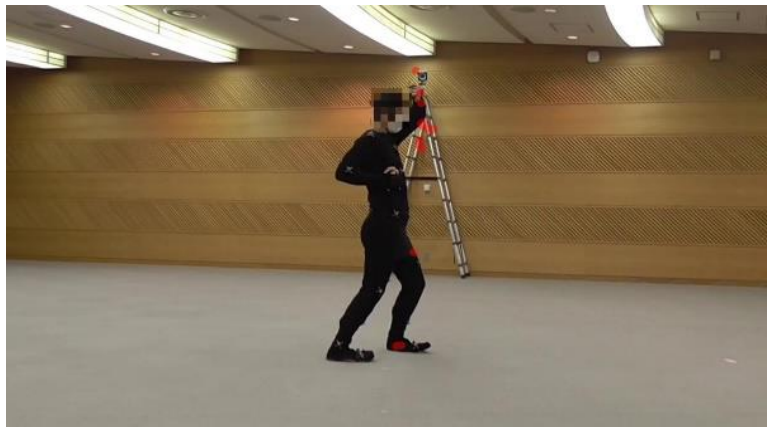
Figure 5 Scene from gaze instruction slideshow

For the annotation group, the slideshow was created based on the expert's verbal comments to ensure consistency with the gaze group. The slideshow, divided into 21 action units, included annotations on the still images extracted from the expert's comments, such as posture, hand position, and limb tension, presented in bullet points to highlight areas of interest (Fig. 6). Each image was displayed for five seconds to allow enough time to read the text. Annotations generally used the expert's original wording, but when the expert pointed with a mouse during the verbal comments to specify left or right, this information was incorporated into the annotations. Additionally, the term *Shime* used in the comments was rephrased as tension to make it easier for non-experts to understand. These annotations avoided indicating the quality of the movement and focused solely on highlighting the areas of interest. Furthermore, following a previous study that suggested the necessity of focusing on multiple areas[16] and the expert's comments in this study often addressed multiple body parts within a single action, multiple annotations were displayed on some still images.

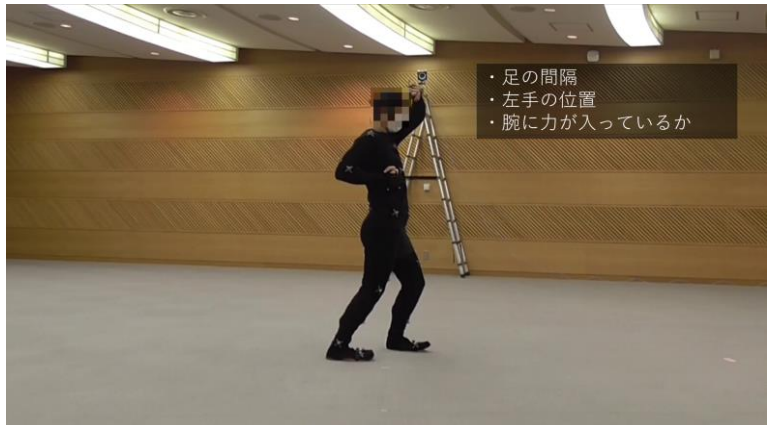

Figure 6 Scene from annotation instruction slideshow
Annotations (in Japanese): (Top) Spacing between feet,
(Middle) Position of left hand, (Bottom) Whether arm tensed

### 2.1.5 Experimental Procedure (An expert)

The outline of the observation experiment conducted with an expert practitioner is shown in Fig. 7. The basic experimental procedure was the same as described for non-expert participants in Section 2.1.2. However, instead of the instruction and second observation round, the expert was asked to observe a video that combined footage from an eye-tracking camera with his own gaze positions during video observation superimposed. While watching the video, the expert could pause at any point and provide verbal comments on aspects that influenced his evaluations during ranking, including details related to skill levels. This procedure is similar to the one described in the previous study[12], and it is expected to reveal the expert's focal points and interpretations. The expert did not report any subjective discrepancies regarding the measured gaze positions, suggesting the reliability of the gaze data.

The ranking results for four performers were obtained from the expert. In response to the question, "Is there a possibility that rankings might differ depending on the evaluator?" the expert answered, "It is highly unlikely." This is because the evaluation criteria and focal points for judging and coaching are commonly understood among instructors, and the parts of the body to observe for each movement are agreed upon. For these reasons, the rankings determined by the expert in this study are used as the correct rankings.

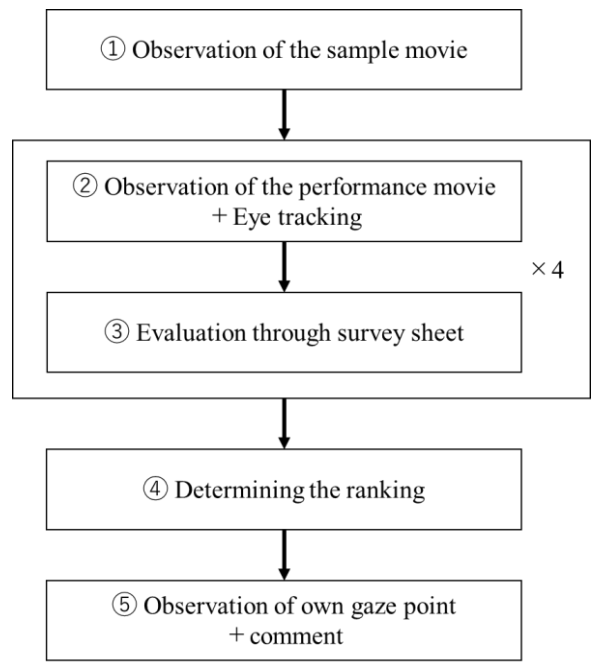

Figure 7 Protocol of experiment for expert

## 2.2 Analysis Methods

### 2.2.1 Analysis of Gaze Positions

The outline of the gaze position analysis method is shown in Fig. 8. In conventional methods, an Area of Interest (AOI) is defined on snapshots (still images extracted from specific frames of a video) of the observation target, and it is determined whether the gaze was directed at the AOI. This method is considered simple and effective when observing stationary targets or targets with minimal movement. However, when the observation target is in motion, it becomes necessary to divide the video into small time intervals and define AOIs for each frame. Alternatively, to ensure that the moving target remains within the AOI, the AOI must be defined broadly, which results in reduced precision due to the ambiguity of the AOI boundaries. Thus, accurately determining gaze positions on moving observation targets using conventional methods requires significant effort.

This study uses a machine learning model called BodyPix to estimate the body parts in the video. The estimation results are then compared with gaze position data obtained from the EMR by applying a projective transformation. This approach enables a high-precision analysis of which body parts of the performer the participants were observing.

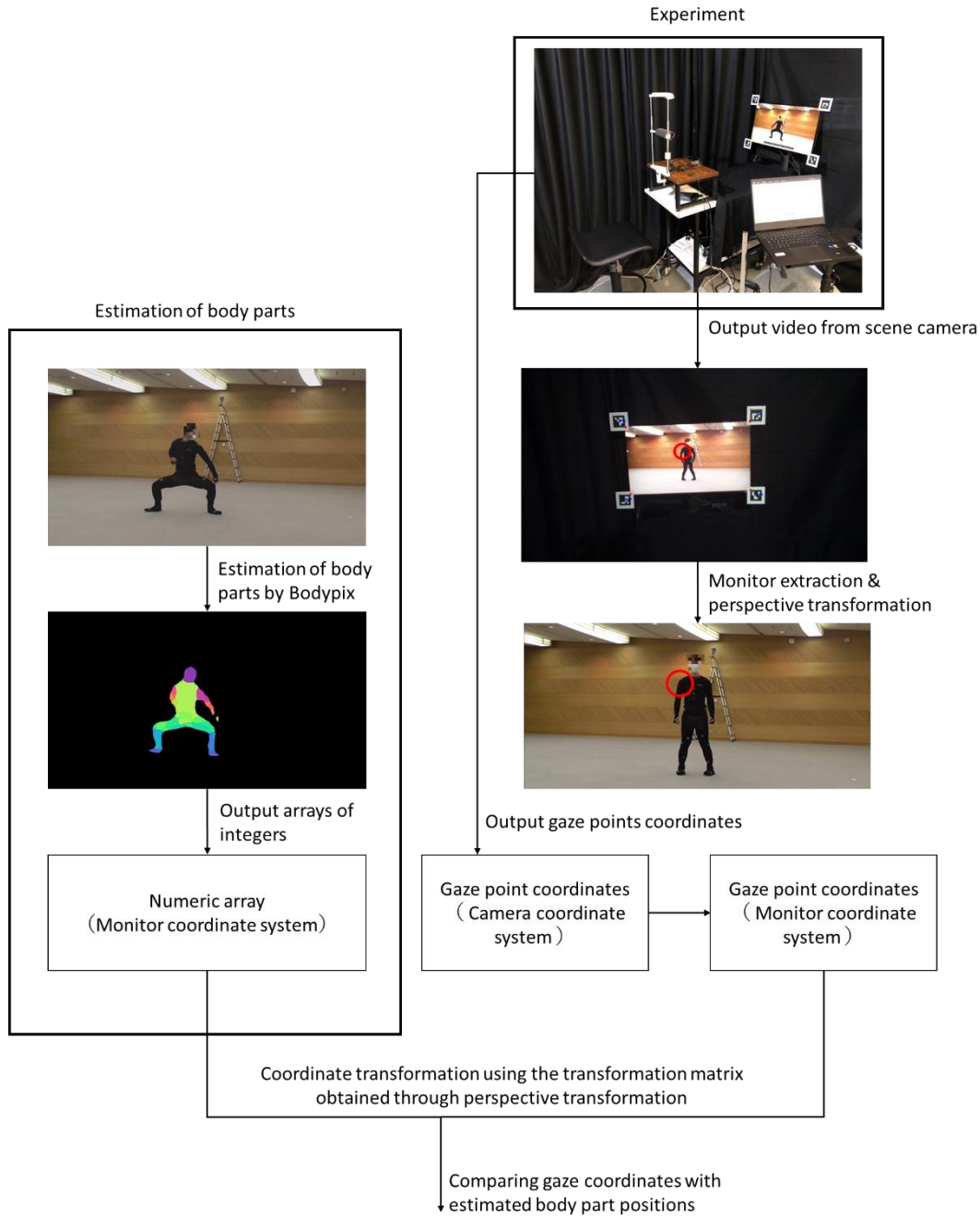

Figure 8 Outline of the method for determining the location of the gaze points

First, the estimation of body parts is explained below. In this study, BodyPix estimates and segments the human body parts to determine which parts the participants observe in the performance videos based on gaze data. BodyPix is an open-source machine-learning model capable of segmenting (classifying) areas of body parts at the pixel level[22]. Fig. 9 shows an example of a performance video segmented by BodyPix, where each body part is color-coded for visualization.

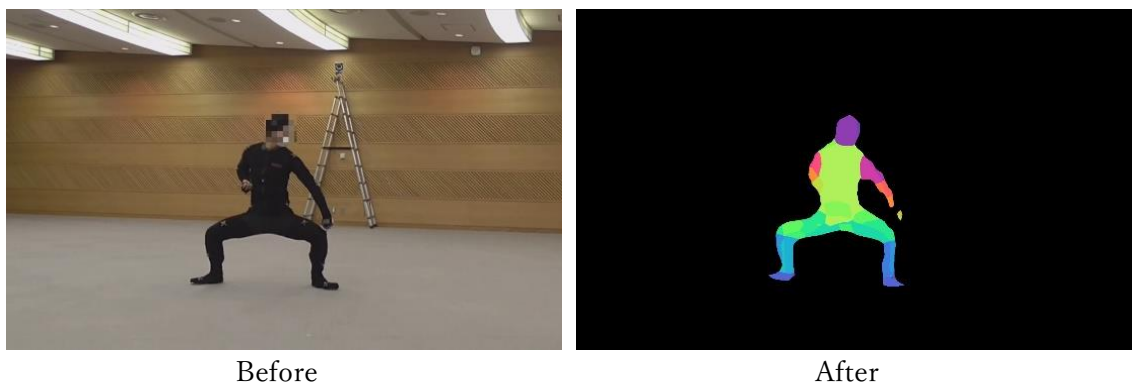
Before　　　　　　　　　　　　　　After
Figure 9 Area segmentation

The performance videos were recorded at a resolution of 1920×1080 pixels, and the resolution was used as the basis for the analysis. BodyPix was employed to classify each pixel as a body or a background. If classified as a body, the pixel was further categorized into 24 parts (represented by integers 0 to 23). This process automated the AOI setting for each frame. BodyPix provides the predefined integer labels for each body part, and the model outputs the segmentation results for every frame in the video. Background pixels were assigned a value of -1. The default configuration of BodyPix treats areas outside the body as a single area for analysis.

However, comments from the expert practitioner highlighted the importance of spatial areas as critical observation points. For example, the expert noted observations such as "checking the tensioning of the lower body by looking at the space between the legs" and "if the elbow position opens up during the front block, it cannot effectively block an attack." These comments indicate that spatial areas, such as the space between the torso and arms, between arms or legs, are essential for evaluating tension and movement accuracy.

To address this, the present study introduced a method to enclose both arms or legs within rectangular areas for analysis. These areas were defined based on the upper and lower boundaries and the leftmost and rightmost edges of the arms or legs. Fig. 10A shows the original segmentation of arms and legs by BodyPix, while Fig. 10B illustrates the results of enclosing the arms and legs within rectangular areas. In the figure, the arms are shown in red, and the legs are shown in green.

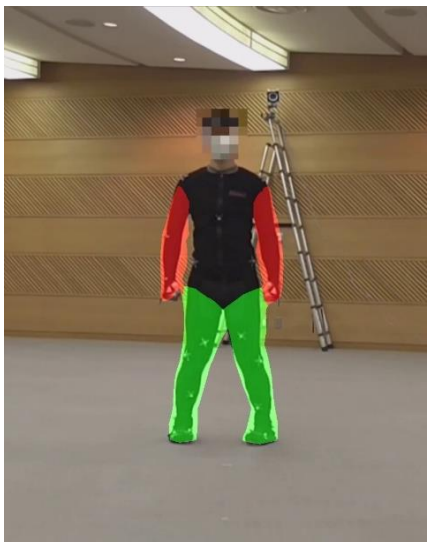
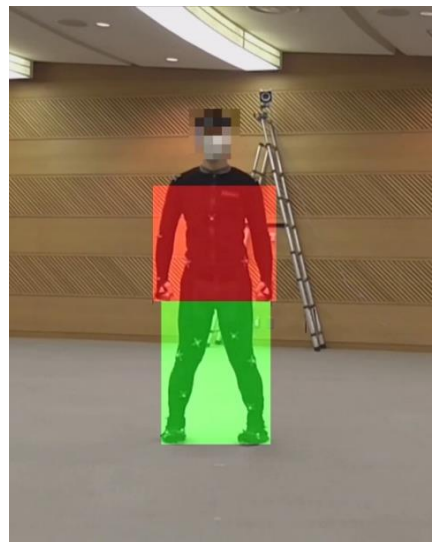

A: Default segmentation

B: Modified segmentation enclosing the arms and legs within rectangular areas

Figure 10 Detection results of upper and lower limb areas by BodyPix

Next, the transformation of gaze coordinate data obtained by the EMR is explained below. The EMR records gaze position coordinates on a frame-by-frame basis, but these coordinates are based on the camera's image resolution corresponding to the eye-tracking camera's footage. Therefore, to accurately determine the gaze position within the actual frame of the performance videos, it is necessary to transform the coordinates into a coordinate system corresponding to the videos displayed on the monitor. Following the method described in the previous study[16], the eye-tracking footage was converted into still images for each frame. Homography transformation was applied to map the monitor area onto an image with the same resolution as the performance videos (1920 pixels wide and 1080 pixels high). For simplicity, the coordinate system corresponding to the eye-tracking camera footage is referred to as the camera coordinate system, while the coordinate system corresponding to the video displayed on the monitor is referred to as the monitor coordinate system.

To extract the monitor in the eye-tracking camera's footage, AR markers generated using ArUco, an extension module of OpenCV, were utilized. The AR markers were affixed to the four corners of the monitor using double-sided tape. The coordinates of the four corners of the monitor in the camera coordinate system were obtained by reading these AR markers from the eye-tracking footage. An example of these extracted coordinates is shown in Fig. 11. Note that the extracted coordinates correspond not to the center of the AR markers but to the positions of the monitor's four corners, as indicated by the red dots in Fig. 11.

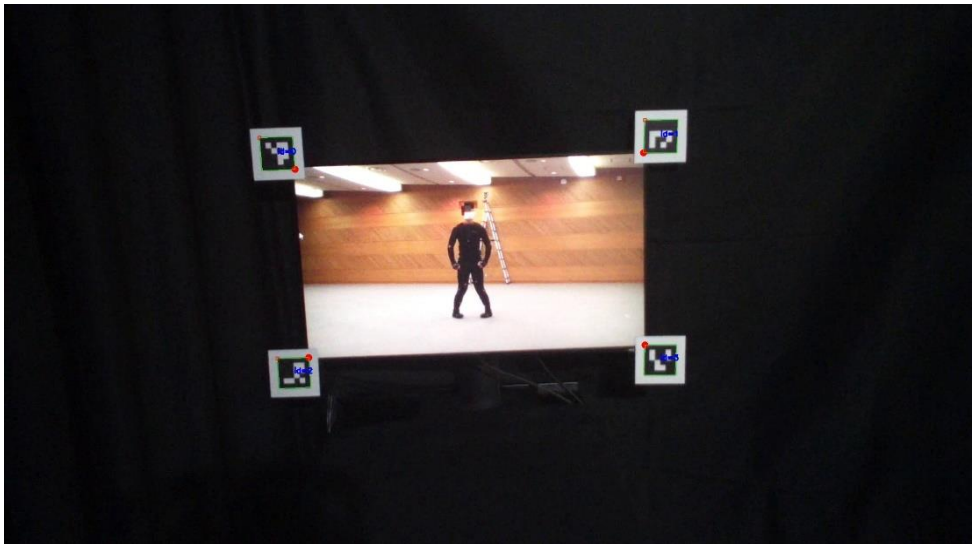

Figure 101 AR markers' position and detection results

Let the coordinates obtained here be represented as $(x_0, y_0)$, $(x_1, y_1)$, $(x_2, y_2)$, $(x_3, y_3)$. For each frame of the video, a homography transformation matrix is calculated to map these coordinates to the monitor coordinate system. The homography transformation matrix $H$ is expressed as follows, and the transformation of the camera coordinates $(x, y)$ to the monitor coordinates $(x', y')$ is considered. The following Equation can represent this transformation:

$$\begin{bmatrix} X' \\ Y' \\ W' \end{bmatrix} = H' \begin{bmatrix} x \\ y \\ 1 \end{bmatrix} = \begin{bmatrix} h'_{00} & h'_{01} & h'_{02} \\ h'_{10} & h'_{11} & h'_{12} \\ h'_{20} & h'_{21} & h'_{22} \end{bmatrix}$$

$$\begin{bmatrix} x' \\ y' \end{bmatrix} = \begin{bmatrix} X'/W' \\ Y'/W' \end{bmatrix}$$

Here, considering $H'$, which is a scaled version of $sH'$, the following Equation is obtained:

$$sH' \begin{bmatrix} x \\ y \\ 1 \end{bmatrix} = \begin{bmatrix} sX' \\ sY' \\ sW' \end{bmatrix}$$

$$\begin{bmatrix} x' \\ y' \end{bmatrix} = \begin{bmatrix} sX'/sW' \\ sY'/sW' \end{bmatrix} = \begin{bmatrix} X'/W' \\ Y'/W' \end{bmatrix}$$

Therefore, the homography transformation matrix is invariant to scaling. By setting $h_{22} = 1$,

$$H = \begin{bmatrix} h_{00} & h_{01} & h_{02} \\ h_{10} & h_{11} & h_{12} \\ h_{20} & h_{21} & 1 \end{bmatrix} = \frac{1}{h'_{22}} \begin{bmatrix} h'_{00} & h'_{01} & h'_{02} \\ h'_{10} & h'_{11} & h'_{12} \\ h'_{20} & h'_{21} & h'_{22} \end{bmatrix}$$

and expanding the transformation equation from $x_0$ to $x'_0$, the following Equation is obtained:

$$x'_0 = H x_0$$

$$\Leftrightarrow \begin{bmatrix} x'_0 \\ y'_0 \\ 1 \end{bmatrix} = \frac{1}{W'_0} \begin{bmatrix} X'_0 \\ Y'_0 \\ W'_0 \end{bmatrix} = \frac{1}{h_{20}x_0 + h_{21}y_0 + 1} \begin{bmatrix} h_{00}x_0 + h_{01}y_0 + h_{02} \\ h_{10}x_0 + h_{11}y_0 + h_{12} \\ h_{20}x_0 + h_{21}y_0 + h_{22} \end{bmatrix}$$

$$\Leftrightarrow (h_{20}x_0 + h_{21}y_0 + 1) \begin{bmatrix} x'_0 \\ y'_0 \\ 1 \end{bmatrix} = \begin{bmatrix} h_{00}x_0 + h_{01}y_0 + h_{02} \\ h_{10}x_0 + h_{11}y_0 + h_{12} \\ h_{20}x_0 + h_{21}y_0 + h_{22} \end{bmatrix}$$

$$\Leftrightarrow \begin{bmatrix} h_{20}x_0 x'_0 + h_{21}y_0 x'_0 + x'_0 \\ h_{20}x_0 y'_0 + h_{21}y_0 y'_0 + y'_0 \end{bmatrix} = \begin{bmatrix} h_{00}x_0 + h_{01}y_0 + h_{02} \\ h_{10}x_0 + h_{11}y_0 + h_{12} \end{bmatrix}$$

$$\Leftrightarrow \begin{bmatrix} x'_0 \\ y'_0 \end{bmatrix} = \begin{bmatrix} h_{00}x_0 + h_{01}y_0 + h_{02} - h_{20}x_0 x'_0 - h_{21}y_0 x'_0 \\ h_{10}x_0 + h_{11}y_0 + h_{12} - h_{20}x_0 y'_0 - h_{21}y_0 y'_0 \end{bmatrix}$$

$$\Leftrightarrow \begin{bmatrix} x'_0 \\ y'_0 \end{bmatrix} = \begin{bmatrix} x_0 & y_0 & 1 & 0 & 0 & 0 & -x_0 x'_0 & -y_0 x'_0 \\ 0 & 0 & 0 & x_0 & y_0 & 1 & -x_0 y'_0 & -y_0 y'_0 \end{bmatrix} \begin{bmatrix} h_{00} \\ h_{01} \\ h_{02} \\ h_{10} \\ h_{11} \\ h_{12} \\ h_{20} \\ h_{21} \end{bmatrix} \quad (1)$$

The same procedure is applied to the other three points and the system of equations in Equation (1) is solved. Here, the coordinates of the four points before the transformation are substituted with the coordinates obtained from the AR markers, while the coordinates of the four points after the transformation are set as follows: $(x'_0, y'_0) =$

$(0,0)$, $(x'_1, y'_1) = (1920, 0)$, $(x'_2, y'_2) = (0, 1080)$, $(x'_3, y'_3) = (1920, 1080)$. Using this transformation matrix, the gaze coordinates in the camera coordinate system calculated by the EMR are transformed into the monitor coordinate system. Finally, by comparing the estimated body part data with the gaze coordinates obtained by the EMR, the body parts of the performer that the participants were observing are analyzed. Following the above procedure reduced the time required for gaze position analysis and simplified the process.

### 2.2.2 Fixation Determination

Fixation determination was performed for the transformed gaze coordinates described in Section 2.2.1. Measured eye movements can be broadly categorized into fixations and saccades. Human perception is achieved by alternately repeating fixations and saccades; however, visual information obtained during saccades is imprecise, and most of the visual information is acquired during fixations. Therefore, this study focuses on fixation components for analysis.

The definition of fixation varies among studies and is often determined based on the duration of the gaze, the angular velocity of eye movements, or a combination of these factors[23]. Additionally, the state of fixation depends on the characteristics of the presented observation task or the observed target, such as whether the target is moving, its size, or positional changes. Thus, it is necessary to consider the appropriate definition of fixation according to the observation conditions. In this study, we adopted the definition proposed by Fukuda et al., where fixation is defined as a state in which the angular velocity of eye movement is approximately 11 deg/sec or less and continues for 165 ms or more when observing a moving target[24]. This definition was adjusted for the experimental conditions in this study (monitor aspect ratio: 16:9, monitor resolution: 1920×1080 pixels, distance between the monitor and the participants: 1 m, video frame rate: 25 fps). As a result, fixation was defined as a state where the gaze movement speed is 600 pixels/sec and less on the monitor resolution and continues for 160 ms or more.

### 2.2.3 Spatiotemporal Characteristics of Eye Movements

The evaluation metrics used in this study were total fixation duration, fixation count, and fixation area. The total fixation duration was calculated as the sum of the durations of each fixation, expressed in seconds. The fixation count was determined based on the definition provided in Section 2.2.2, counting instances where the gaze was maintained for 160 ms or longer. Additionally, following the previous study{Sasamoto, 2020 #2} by Sasamoto et al., fixations were classified into instantaneous fixations (duration less than 500 ms) and deliberate fixations (duration of 500 ms or more), and the counts for each category were recorded. The total fixation count was defined as the summation of these two categories.

Two metrics were used to evaluate fixation areas: the total number of fixation areas and the fixation proportion by area. The total number of fixation areas was calculated using the BodyPix segmentation results, which categorized the body into 11 areas described in Section 2.2.1. Specifically, the number of areas where fixations occurred was counted for each video frame. Subsequently, the data for the three segments of interest (the three movements shown in Fig. 4, selected from the 21 movements described in Section 2.1.4) were aggregated to define the total number of fixation areas. The fixation proportion by area was calculated based on the frames in which fixations occurred. Specifically, for the three segments shown in Fig. 4, the total number of fixation frames was divided by the total number of frames in the three segments to compute the fixation proportion.

### 2.2.4 Evaluation of Ranking Results

To evaluate the ranking results, the score $S_j$ was used, which represents the total distance between the ranking $G_i$ determined by the expert for performance video $i$ and the ranking $g_j$ determined by the non-expert participants. A smaller $S_j$ value indicates that the rankings determined by the non-expert participants are closer to those of the expert. A smaller $S_j$ value in the second round compared to the first round, reflecting improved performance in ranking determination. The formula for calculating $S_j$ is shown in Equation (2). Here, $j$ represents the non-expert participants index. In this experiment, ties were not allowed, so the possible values of $S_j$ were 0, 2, 4, 6, and 8.

$$S_j = \sum_{i=1}^{4} |G_i - g_i| \qquad (2)$$

**2.2.5 Statistical Analysis**

In this experiment, standard deviations due to individual differences were large. Cohen's d was used as an indicator of effect size to compare the results before and after instruction. Cohen's d is calculated by dividing the mean difference between two groups by their standard deviation. The interpretation of Cohen's d follows the standard guidelines: 0.2 indicates a small effect, 0.5 indicates a medium effect, and 0.8 indicates a large effect[25, 26].

**3. Results of the Experiment with Non-experts**

In this study, we conducted an observational experiment using four performance videos, focusing on the video ranked first by the expert. Each video was divided into 21 movements, and an evaluation was conducted by combining three distinctive segments of interest (*Jodan uke, Jodan tsuki,* and *Shomen uke*) highlighted in Fig. 4. Based on a post-experiment interview with the non-experts, we excluded two participants who intentionally ignored the instruction points and one participant who determined the ranking based on bias (subjectively assuming that the last video shown was the best). The results are presented for the remaining 25 participants.

**3.1 Eye Tracking Results**

Representative non-expert participants were selected for each group, where distinctive tendencies were observed, and their gaze positions before and after the instructions were plotted (Fig.12). As shown in Fig.5, the gaze positions of the expert participant were concentrated mainly on the left arm, torso, and the space between them, with some focus also on the left leg. In contrast, in Fig.12A, the non-expert participant in the gaze group initially focused on the torso. However, after the instructions, he/she also directed his/her attention to the arms and legs. Additionally, as described in Fig.6, the expert commented that he observed "the spacing of the feet," "the position of the left hand," and "whether there was tension in the arms." The non-expert in the annotation group (Fig.12B) initially distributed his/her gaze across the entire body, but after the instructions, his/her gaze became more focused on areas such as the left arm, right hand, and legs. On the other hand, the non-experts in the control group, who did not receive instructions, showed little change in his/her gaze positions (Fig.12C).

Note, in Fig.12, the gaze positions during fixation and saccade states for one of the distinctive movements, *Jodan uke,* are plotted as red dots. It was confirmed that the tendencies remained unchanged even when plotting only on fixations. We divided eye movements into fixation and saccades based on the definitions described in Section 2.2.2 to analyze the spatiotemporal changes in eye movement. We analyzed three parameters: fixation duration, fixation counts, and fixation areas. In this study, fixation was defined as lasting 160 ms or more, although it has been reported in other studies that durations of 100 ms or more are often used[23]. Upon reviewing the data obtained in this experiment, fixations lasting 80 ms were also observed. Supplementary Material Section 1 provides an analysis where fixation is defined as an eye movement speed of 600 pixels/sec and less sustained for 80 ms or more.

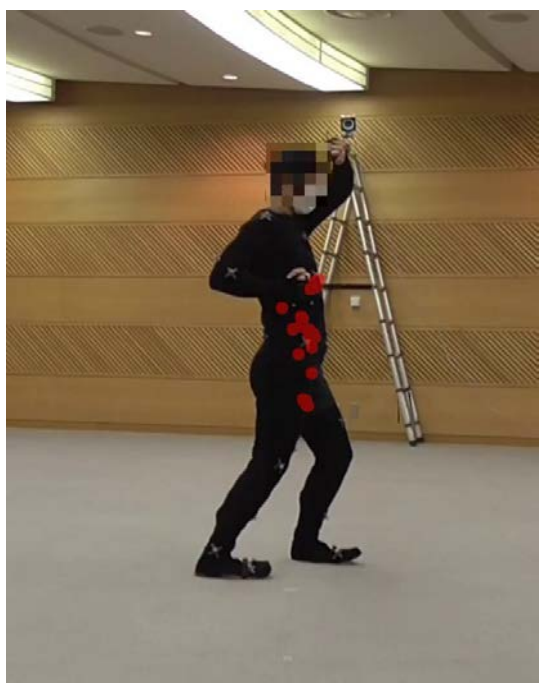
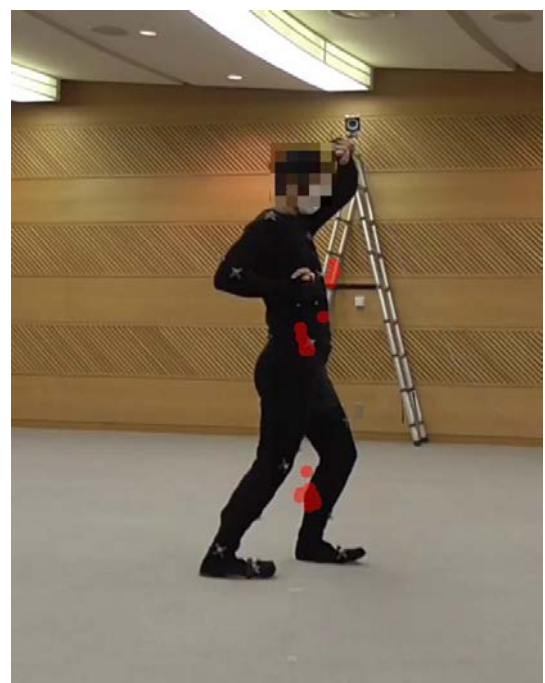

Before　　　　　　　　　　　　　　　After

A: Gaze group

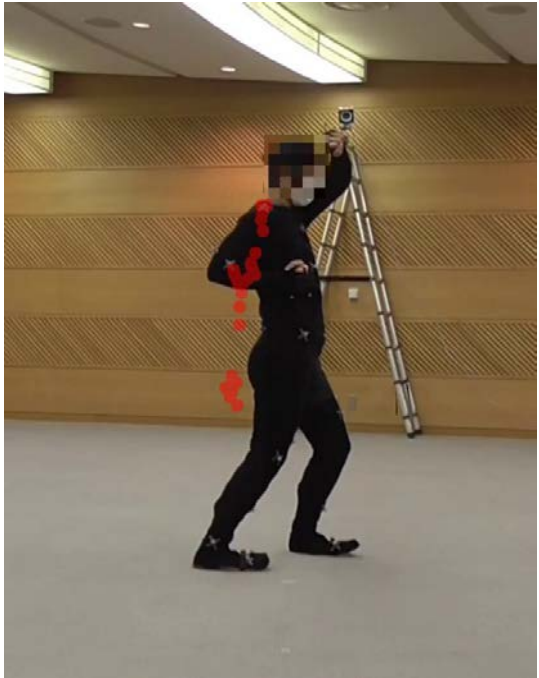
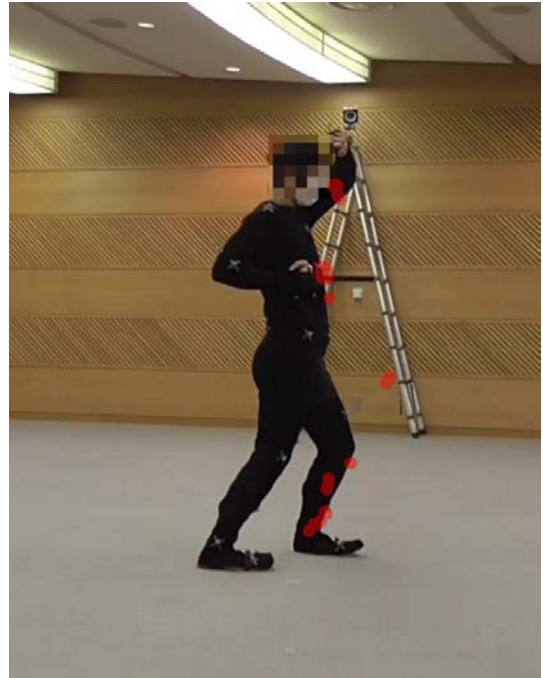

Before　　　　　　　　　　　　　　After

B: Annotation group

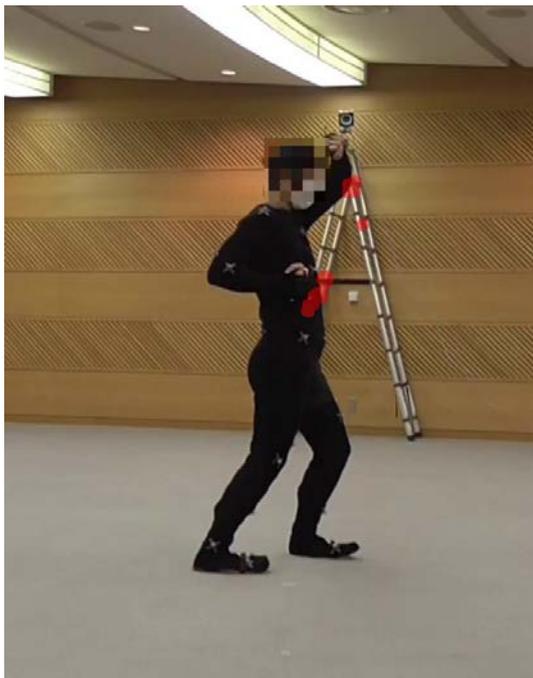
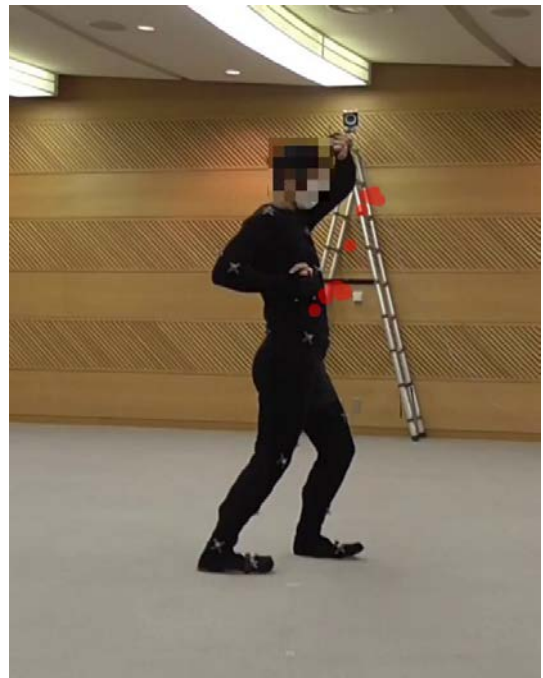

Before　　　　　　　　　　　　　　After

C: Control group

Figure 12 Feature of gaze patterns (Respective subject in each group)

Representative datasets were selected for each group, where distinctive tendencies were observed, and the number of fixations by fixation duration before and after the instructions was plotted (Fig.13). Red dashed lines indicate the thresholds for instantaneous and deliberate fixations.

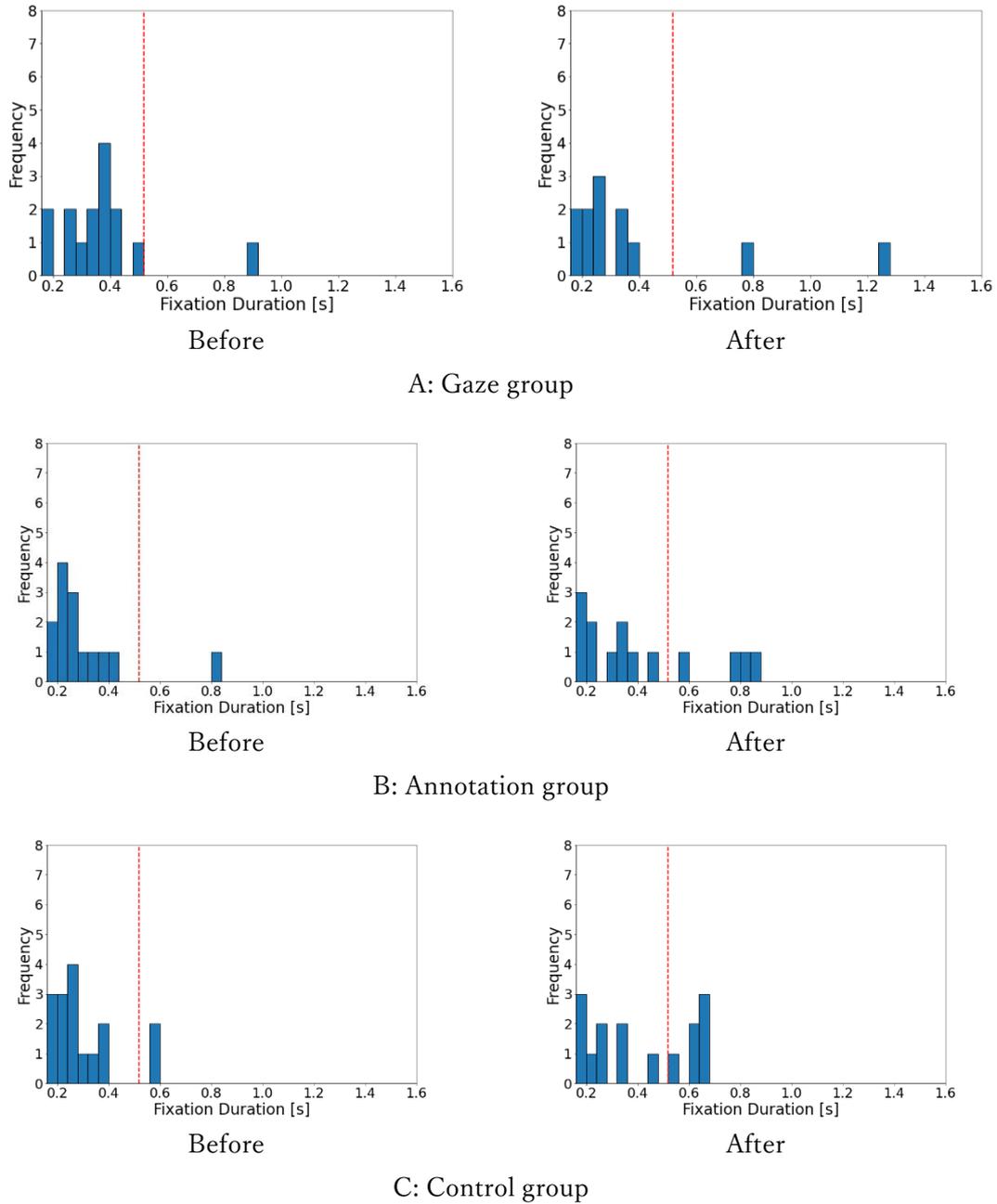

Before / After

A: Gaze group

Before / After

B: Annotation group

Before / After

C: Control group

Figure 13 Fixation frequencies categorized by duration (Respective subject in each group)

For all groups, the frequency of instantaneous fixations was higher both before and after the instructions. While the gaze group showed little change in tendencies after the instructions, the annotation and control groups exhibited an increase in the frequency of deliberate fixations. This tendency was slightly more pronounced in the control group.

### 3.1.1 Fixation duration

Table 1 and Table S.1 present the measured values of the total fixation duration of each group before and after the instructions and that of the expert. Consistently across both tables, a small decrease trend in total fixation duration was observed in the gaze group, no significant change was observed in the annotation group, and a small increase trend was observed in the control group, provided that Cohen's d was 0.2 or greater, indicating a small effect size. No significant difference was observed between non-experts and the expert in the total fixation duration. Approximately 60% of the duration was judged as fixation, compared to the total duration for the three segments of interest (7.6 seconds), and the remaining 40% of the duration corresponded to saccades.

Table 1 Total Fixation Duration [s]

| Expert | Gaze | | Annotation | | Control | |
|---|---|---|---|---|---|---|
| 4.36 | Before | After | Before | After | Before | After |
| Mean±S.D. | 4.22±1.87 | 3.73±2.33 | 4.58±1.97 | 4.35±1.73 | 4.14±2.08 | 4.63±1.90 |
| Cohens' d | -0.24 | | -0.12 | | 0.24 | |

### 3.1.2 Fixation counts

This study investigated the increase or decrease in instantaneous and deliberate fixations mentioned in a previous study[15]. Following the thresholds used in the previous study, instantaneous fixations were defined as fixations with a duration of less than 500 ms, and deliberate fixations were defined as fixations with a duration of 500 ms or more. Table 2 and Table S.2 present the measured values for the number of instantaneous fixations, deliberate fixations, and their summation (i.e., total fixation count) before and after the instructions for each group, as well as the expert.

Consistently across both tables, no significant change in the number of instantaneous fixations, deliberate fixations, or total fixation counts was observed in the gaze group. In the annotation group, consistently across both tables, no significant change was observed in the number of instantaneous fixations or total fixation counts, while a small increasing trend was observed in the number of deliberate fixations. In the control group, consistently across both tables, a small decreasing trend was observed in the number of instantaneous fixations, and a small increasing trend was observed in the number of deliberate fixations. However, for total fixation counts, no significant change was observed for fixations lasting 160 ms or more (Table 2), while a small decreasing trend was observed for fixations lasting 80 ms or more (Table S.2).

Table 2 Fixation counts [-]

|  | Instantaneous | | Deliberate | | Summation | |
|---|---|---|---|---|---|---|
| Expert | 12 | | 2 | | 14 | |
|  | Before | After | Before | After | Before | After |
| Gaze | 9.00±4.65 | 8.43±3.82 | 1.71±1.11 | 1.57±1.27 | 10.86±4.98 | 10.00±4.62 |
|  | -0.13 | | -0.12 | | -0.18 | |
| Annotation | 9.38±5.26 | 8.75±3.54 | 2.25±2.25 | 2.88±1.81 | 11.63±4.81 | 11.63±4.14 |
|  | -0.14 | | 0.31 | | 0.00 | |
| Control | 8.70±4.35 | 7.30±2.79 | 2.40±1.84 | 3.40±2.46 | 11.10±5.04 | 10.70±3.09 |
|  | -0.38 | | 0.46 | | -0.10 | |

### 3.1.3 Fixation Areas

Table 3 and Table S.3 present the gaze measurement results for the expert and each group before and after the instructions. These include the total number of fixation areas and the fixation proportions for the upper and lower limbs, face/torso, background (outside the performer's body visible in the video), and saccades, excluding the space between both arms and legs. Consistently across both tables, the gaze group exhibited a large decreasing trend in the total number of fixation areas. Specifically, a large decreasing trend was observed in the proportion of upper and lower limbs, while no significant trends were observed for the face/torso or the background.

In the annotation group, a small increasing trend was observed in the total number of fixation areas. Consistently across both tables, the proportion of upper and lower limbs showed a small increasing trend, while the proportion of the face/torso showed a small decreasing trend. The proportion of the background showed no significant change for fixations lasting 160 ms or more (Table 3), but an increasing trend was observed for fixations lasting 80 ms or more (Table S.3).

In the control group, consistently across both tables, a small increasing trend was observed in the total number of fixation areas. Specifically, a small increasing trend was observed in the proportion of upper and lower limbs as well as the background, while no significant trends were observed for the face/torso.

Table 3 Total number of fixation areas and proportion by area
(excluding the space between both arms and legs)

|  | Total number of fixation areas | | Proportion Arms and Legs [%] | | Proportion Face and Torso [%] | | Proportion Outside of body [%] | | Proportion Saccard [%] | |
|---|---|---|---|---|---|---|---|---|---|---|
| Expert | 7 | | 18.42 | | 18.42 | | 23.16 | | 40.00 | |
|  | Before | After | Before | After | Before | After | Before | After | Before | After |
| Gaze | 6.29 ±1.70 | 4.71 ±1.11 | 17.59 ±7.87 | 11.05 ±8.62 | 25.26 ±18.98 | 23.16 ±20.77 | 13.38 ±10.93 | 15.41 ±10.99 | 43.76 ±24.96 | 50.38 ±30.50 |
|  | -1.09 | | -0.79 | | -0.11 | | 0.19 | | 0.24 | |
| Annotation | 5.25 ±2.12 | 6.25 ±2.71 | 12.70 ±5.78 | 16.78 ±11.62 | 32.89 ±19.08 | 23.95 ±19.10 | 15.59 ±11.55 | 17.63 ±11.87 | 38.82 ±26.17 | 41.64 ±23.59 |
|  | 0.41 | | 0.44 | | -0.47 | | 0.17 | | 0.11 | |
| Control | 4.80 ±1.87 | 5.30 ±1.34 | 7.16 ±9.45 | 10.89 ±8.89 | 33.74 ±23.47 | 32.21 ±19.03 | 14.74 ±17.09 | 18.47 ±14.40 | 44.37 ±27.89 | 38.42 ±25.48 |
|  | 0.31 | | 0.41 | | -0.07 | | 0.24 | | -0.22 | |

Table 4 and Table S.4 present the gaze measurement results for the expert and each group before and after the instructions. These include the fixation proportions for the upper and lower limbs (treated as a single rectangular area, including the space between both arms and legs), face/torso, background (outside the performer's body), and saccades. In the gaze group, for fixations lasting 160 ms or more (Table 4), no significant trends were observed in the proportions of the upper and lower limbs or face/torso. However, a medium decreasing trend was observed in the proportion of the background (Cohen's d was greater than 0.5, indicating a medium effect size). For fixations lasting 80 ms or more (Table S.4), small decreasing trends were observed in the proportions of the upper and lower limbs and the background, while a small increasing trend was observed in the proportion of the face/torso.

In the annotation group, consistently across both tables, no significant trends were observed in the proportion of the upper and lower limbs, while a small decreasing trend was observed in the proportion of the face/torso. The proportion of the background showed a small increasing trend for fixations lasting 160 ms or more (Table 4).

In the control group, consistently across both tables, a small increase trend in the proportion of the upper and lower limbs was observed. For the face/torso, a small decreasing trend was observed for fixations lasting 160 ms or more (Table 4), while no significant changes were observed for fixations lasting 80 ms or more (Table S.4). For the background, no significant trends were observed for fixations lasting 160 ms or more (Table 4), but a small increasing trend was observed for fixations lasting 80 ms or more (Table S.4).

Table 4 Proportion by area (including the space between both arms and legs)

|  | Proportion Arms and Legs [%] | | Proportion Face and Torso [%] | | Proportion Outside of body [%] | | Proportion Saccard [%] | |
|---|---|---|---|---|---|---|---|---|
| Expert | 42.63 | | 17.37 | | 16.84 | | 40.00 | |
|  | Before | After | Before | After | Before | After | Before | After |
| Gaze | 41.73 ±19.65 | 32.56 ±19.98 | 7.74 ±7.65 | 12.41 ±15.27 | 6.77 ±8.64 | 4.66 ±5.16 | 43.76 ±24.96 | 50.38 ±30.50 |
|  | 0.08 | | 0.06 | | -0.50 | | 0.24 | |
| Annotation | 40.07 ±16.44 | 40.99 ±21.57 | 13.09 ±12.54 | 8.29 ±11.74 | 8.03 ±8.15 | 9.08 ±9.91 | 38.82 ±26.17 | 41.64 ±23.59 |
|  | 0.02 | | -0.23 | | 0.33 | | 0.11 | |
| Control | 35.58 ±26.32 | 42.26 ±22.58 | 9.58 ±12.49 | 5.11 ±4.62 | 10.47 ±13.59 | 14.21 ±12.06 | 44.37 ±27.89 | 38.42 ±25.48 |
|  | 0.30 | | -0.26 | | 0.03 | | -0.22 | |

For comparison, Table 5 and Table S.5 show the fixation proportions for the space of both arms and legs, which was included by expanding the upper and lower limb area to encompass the rectangular area between both arms and legs. This proportion was calculated by subtracting the combined fixation proportion for the upper and lower limbs and the face/torso defined in Table 4 and Table S.4 from the combined fixation proportion for the upper and lower limbs and the face/torso defined in Table 3 and Table S.3.

Table 5 Proportion of space between both arms and both legs [%]

|  | Gaze | | Annotation | | Control | |
|---|---|---|---|---|---|---|
|  | Before | After | Before | After | Before | After |
| Mean±S.D. | 6.62±5.20 | 10.75±10.10 | 7.57±6.78 | 8.55±5.85 | 4.26±4.91 | 4.26±3.33 |
| Cohens' d | 0.51 | | 0.16 | | 0.00 | |

In the gaze group, consistently across both tables, a medium increasing trend was observed in the proportion of the space between both arms and legs, while in the control group, no significant changes were observed. In the annotation group, no significant changes were observed for fixations lasting 160 ms or more (Table 5), whereas a small increasing trend was observed for fixations lasting 80 ms or more (Table S.5).

### 3.2 Ranking Results

Table 6 shows the ranking scores of non-experts before and after the observation with instructions. A small increase trend in scores was observed across all groups: the gaze, annotation, and control groups. In the gaze group, out of 7 participants, 5 showed score improvement, 0 showed no change, and 2 showed a decline. In the annotation group, out of 8 participants, 3 showed score improvement, 3 showed no change, and 2 showed a decline. In the control group, out of 10 participants, 3 showed score improvement, 6 showed no change, and 1 showed a decline. Based on these results, the proportion of participants with score improvement was 71.4% in the gaze group, 37.5% in the annotation group, and 30% in the control group.

Table 6 Scores for the ranking determination of each subject

|  | Gaze | | Annotation | | Control | |
| --- | --- | --- | --- | --- | --- | --- |
|  | before | after | before | after | before | after |
| S1 | 4 | 2 | 8 | 6 | 4 | 4 |
| S2 | 2 | 0 | 4 | 2 | 6 | 0 |
| S3 | 2 | 4 | 2 | 2 | 2 | 2 |
| S4 | 4 | 2 | 4 | 4 | 2 | 2 |
| S5 | 2 | 0 | 2 | 2 | 4 | 4 |
| S6 | 4 | 2 | 0 | 2 | 4 | 4 |
| S7 | 0 | 4 | 4 | 2 | 2 | 2 |
| S8 |  |  | 4 | 4 | 0 | 4 |
| S9 |  |  |  |  | 6 | 4 |
| S10 |  |  |  |  | 4 | 0 |
| Mean±S.D. | 2.6±1.5 | 2.0±1.6 | 3.5±2.3 | 3.0±1.5 | 3.4±1.9 | 2.6±1.7 |
| Cohen's d | 0.36 | | 0.25 | | 0.45 | |

Among the four performance videos, only 3 out of 25 participants assigned a rank other than fourth to the video ranked fourth by the expert. This indicates that nearly all participants could identify this video as the poorest performance, suggesting that the difficulty of ranking was low. Therefore, an additional analysis excluding the fourth-ranked video was conducted (Table S.6). When the fourth-ranked video was excluded, compared to the analysis without exclusion, one participant in each of the control and annotation groups shifted from showing improvement to no change in their scores, while no change was observed in the gaze group. According to Table S.6, the control group showed a small increasing trend in scores, whereas the annotation group showed almost no improvement trend. Additionally, the proportion of participants who improved their scores was 25% in the annotation group and 20% in the control group, indicating a reduction in the proportion of improvement.

Further, another additional analysis was conducted excluding the three participants who ranked the fourth-ranked video differently from the expert (Table S.7). In this case, the annotation group showed a small increasing trend in scores, whereas the control group showed almost no improvement trend. Additionally, the proportion of participants who improved their scores was 28.5% in the annotation group and 0% in the control group. Originally, there was a difference in the number of participants among the groups, but excluding the three participants reduced this difference. Furthermore, although there was no significant difference in pre-instruction scores ($p > 0.05$), 71.4% of the participants in the gaze group were able to improve their scores.

## 4. Discussion

### 4.1 Summary and Discussion of Changes in the Spatiotemporal Characteristics of Eye Movements in Each Group

For the expert, although there was no significant difference in total fixation duration compared to the non-experts, the total fixation counts (particularly instantaneous fixations) and the total number of fixation areas were slightly higher. This suggests that the expert gathered information from various areas by directing short gazes to multiple areas. Therefore, efforts to encourage the observation of various areas through instructions can be considered appropriate for teaching the characteristics of eye movements of the expert. Furthermore, regarding the fixation proportions by area, when the space between the upper and lower limbs was excluded, the results showed that the expert focused equally on the upper and lower limbs, the face, and the torso. However, when the space between the upper and lower limbs was included, the proportion of the upper and lower limbs dominated. Comments from the expert frequently emphasized their focus on specific spatial areas, such as the spaces between the torso and arms, between the arms, and between the legs. This result aligns with prior research[12], which reported that gaze points could be extracted using gaze measurement and interviews, and it confirms that the gaze points of the expert could also be extracted in this study.

When examining changes in the spatiotemporal characteristics of representative subjects in each group before and after instruction, Fig. 12 shows that after instruction, their gaze concentrated on the body parts indicated in the instructions, resembling the findings of a previous study[16]. Additionally, Fig. 13 showed an increase in deliberate fixations during the second observation (after instruction), consistent with the findings of another study[17]. Below, we summarize and discuss the average trends of spatiotemporal characteristic changes in the control, annotation, and gaze groups, respectively.

In the control group, the total fixation duration (Section 3.1.1) increased after instruction. The total number of fixations (Section 3.1.2), defined by a duration of 160 ms or longer, showed almost no change, but when defined as 80 ms or longer, there was a decreasing trend. This indicates a reduction in short fixations lasting between 80 ms and 160 ms. The number of instantaneous fixations showed a decreasing trend, while the number of deliberate fixations showed an increasing trend. This suggests that the increased total fixation duration reported in Section 3.1.1 resulted from an increased trend in deliberate fixations. The total number of fixation areas (Section 3.1.3) showed an increasing trend. The fixation proportion at the upper and lower limbs and outside the body also increased. This suggests that participants directed their gaze toward moving distal body parts, such as the hands and feet, which led their gaze outside the body's range. Comments from the second observation, such as "I looked more closely," support the idea that participants intentionally focused on specific body parts during their observations.

In the annotation group, the total fixation duration (Section 3.1.1), the total number of fixations (Section 3.1.2), and the number of instantaneous fixations showed little change. However, the number of deliberate fixations exhibited an increasing trend. The total number of fixation areas (Section 3.1.3) showed an increasing trend. When the space between the upper and lower limbs was excluded, there was a decreasing trend in the proportions of gazes directed at the face and torso and an increasing trend in the proportion of gazes directed at the upper and lower limbs. When the space between the upper and lower limbs was included, the proportion of gazes directed at the face and torso decreased, while the proportion of gazes directed at the upper and lower limbs remained relatively unchanged. Meanwhile, the proportion of gazes directed outside the body increased. These results suggest that participants shifted their gaze from the face and torso to the upper and lower limbs and the surrounding spaces after instruction. These findings align with the results shown in Table 5 and Table S.5. Moreover, interview results suggested that participants adjusted their perspective to evaluate using the instructed viewpoint, and gaze measurement data suggest that multiple areas were observed as instructed. Therefore, instructions using annotations directly specifying areas of interest effectively encouraged participants to observe multiple instructed locations intentionally.

In the gaze group, a decreasing trend in total fixation duration (Section 3.1.1) was observed, and there was little change in the total number of fixations (Section 3.1.2). Although there were no significant trends in the number of instantaneous or deliberate fixations, the average values decreased, suggesting that participants were unlikely to focus on specific areas intentionally. The areas of interest were indicated using gaze plots, so the total number of fixation areas was expected to increase. However, a large decreasing trend was observed in the total number of fixation areas (Section 3.1.3), suggesting that participants narrowed their range of gaze movement and efficiently acquired information in a shorter amount of time. Therefore, although the instructions attempted to encourage participants to observe multiple areas intentionally, it was concluded that this approach failed to encourage participants to observe multiple areas consciously. Regarding the fixation proportions by area, when the space between the upper and lower limbs was excluded, a large decreasing trend in the proportion of gazes directed at the upper and lower limbs was observed, regardless of whether the fixation duration was defined as 160 ms or 80 ms. There were almost no changes in other areas. When the space between the arms and legs was included, there were differences in degree depending on the definition of fixation duration. However, a decreasing trend in the proportion of gazes directed at the upper and lower limbs was commonly observed, and the proportion of gazes directed outside the body decreased. On the other hand, the proportions of gazes directed at the face and torso increased. Table 5 and Table S.5 show an increasing trend in the proportion of gazes directed at the space between the arms and legs. These results suggest that after the instructions, the gaze point shifted from the areas directly above the upper and lower limbs and the outer space of the body to the space between the arms and legs. Therefore, it was concluded that instructions using gaze plots failed to encourage participants to observe multiple instructed areas intentionally.

Similar spatiotemporal characteristics of eye movements in the gaze group (narrowing their range of gaze movement and possibly acquiring information efficiently) were reported by Kato et al. in their study of batters' visual exploration activities during the preparation time for hitting in baseball[27]. They found that experts moved their gaze over a narrower range than non-experts and argued that this result reflects the use of peripheral vision, which is highly effective for grasping information involving temporal changes and spatial relationships. Similar characteristics have also been reported in *kendo*[28] and boxing[29]. This peripheral vision-based characteristic, as shown in Table 5 and Table S.5, was observed relatively prominently in the gaze group before and after instruction (Cohen's d = 0.51, indicating a medium effect size). While it is impossible to measure peripheral vision directly, no comments indicating specific attention to the space itself were observed in the gaze group's questionnaires or interviews. Since participants did not report focusing exclusively on narrow areas, such as the space between the arms and legs, their focus is unlikely to be confined to narrowly defined areas. Instead, it is considered that they utilized peripheral vision to gather information from a wider range.

When comparing the results of this study to those of prior research[17], which employed an educational program combining three methods—gaze, annotations, and practical instruction—the annotation group and control group produced similar results, while the gaze group showed distinct tendencies. Specifically, in the annotation and control groups, the number of instantaneous fixations decreased while deliberate fixations increased. In contrast, in the gaze group, these numbers either showed no significant change or decreased (Table 2 and Table S.2). However, in prior research[17], it is unclear which instructional method contributed to the increase in deliberate fixations. Within the results of this study, it might suggest that the number of deliberate fixations in the annotation and control groups increased after instruction, surpassing that of the expert.

4.2 Discussion of Ranking Results

According to Section 3.2, in the gaze group, the change in ranking results before and after instruction became closer to the ranking results of the expert, both when the fourth video was included and when it was excluded (Cohen's d = 0.36, indicating a small effect). The scores before instruction in the gaze group were relatively lower than the other groups (mean: 2.6) and decreased further after instruction (mean: 2.0). Here, a lower score indicates closer proximity to the ranking results assigned by the expert. The proportion of participants whose scores improved was 71.44%, suggesting that this type of instruction using gaze could enhance scoring ability.

In the annotation group, all eight participants commented that they changed their perspective to evaluate using the instructed viewpoints. However, the change in ranking results before and after instruction, both when the fourth video was included and when it was excluded, showed the largest deviation from the expert rankings among the three groups. Additionally, the proportion of participants with improved scores was 37.4% when the fourth video was included and 25% when it was excluded, which was lower than in the gaze and control groups. Therefore, this type of instruction using annotations appears to have limited potential to improve scoring ability.

One possible reason for these results is that this experiment did not provide instructions on what is correct. As a result, even though the points of focus were conveyed through the instructions, the participants may not have been able to reflect these in their scores. Some of the expert comments included elements related to the correctness defined by the competition, such as the position of the punch or the stance. For instance, the position of the punch is determined by opponents, such as the position of the opponent's nose or the solar plexus. Even if participants observed the arm as instructed, they might find it difficult to evaluate in a manner similar to the expert. On the other hand, aspects such as the stance or whether the lower body is tensioned are more likely to allow evaluations close to those of the expert based on posture and movements. However, some non-experts commented that "the inward-pointing toes seemed off to me" when observing *Sanchin dachi* (Hourglass stance), where the feet are shoulder-width apart with the toes turned inward. This suggests that the concept of what is good (or correct) was not always clearly conveyed. This inference is also discussed in prior research[30]. Watanuki pointed out that while videos are advantageous for recording and preserving skills, the extent to which skills can be acquired from videos greatly depends on the viewer's knowledge and abilities.

In the control group, no significant changes in scores were expected, but Cohen's d was 0.45 when the fourth video was included and 0.26 when it was excluded, indicating a slight improvement. Additionally, 40% of participants, or 30% when including both improvements and declines in scores, showed some change. These results suggest that even in the control group, there were moderate changes in rankings, possibly due to insufficient elimination of novelty. While this does not directly indicate the impact of novelty on scores, prior research[17] has suggested that novelty may influence eye movement characteristics. Sasamoto et al. noted that a potential issue during gaze measurement is that, since participants had not previously viewed comparison videos, the novelty may have led to an excessive increase in gaze counts before instruction. Based on the findings, in this experiment, all participants (who were not fully familiar with *karate kata*) were shown a sample demonstration video (Fig. 3①) to establish a reference for evaluating the performance videos and to mitigate novelty. In post-experiment interviews, many control group participants commented that their points of focus did not change, or they looked at the same areas but observed them more closely. However, three participants whose scores changed more than the others stated, "I established my own evaluation criteria during the second round." This suggests that the novelty was not fully eliminated for some participants, and their score changes might have influenced the results. Supplementary Material Table S.7 shows the results excluding participants with pre-instruction scores of 6 or 8, which had notably low alignment with expert rankings. In these results, no score changes were observed in the control group (Cohen's d = 0.00).

Another factor that may have contributed to the limited impact on ranking results in this experiment is the small number of performance videos. Additionally, the rankings of one performer were noticeably clear due to poor performance, which likely reduced variability in scores among non-experts. In this experiment, ties were not allowed, and possible scores were limited to five specific values: 0, 2, 4, 6, and 8. Of these, 44% of participants scored 0, indicating identical rankings to those of the experts before instruction, or 2, indicating close alignment. This left little room for improvement in scores. Therefore, even if the scoring ability was improved through instruction, reflecting this improvement in the scores might not have been easy.

### 4.3 Integrated Discussion

This study planned to examine whether encouraging participants to observe multiple body parts intentionally, alter eye movements' spatiotemporal characteristics, and improve evaluation ability by comparing two instructional methods—gaze-based and annotation-based—the differences in their effects on ranking scores and changes in eye movement were observed.

The annotation-based instruction demonstrated the potential effect of guiding participants to observe specified areas of interest intentionally. However, despite the increase in the total number of fixation areas, the proportion of participants whose ranking scores improved was small, suggesting that this approach did not lead to an improvement in scoring ability, specifically in terms of alignment with expert rankings. Therefore, these findings support previous research[16], suggesting that focusing solely on specific points highlighted by instructors is insufficient. Simply increasing the number of areas of interest is likely inadequate for effectively transferring scoring abilities.

The results of the gaze-based instruction revealed distinctive trends that differed from the intended aim of encouraging participants to observe multiple areas. Specifically, while it did not increase the number of fixation areas, characteristics consistent with those described in other studies[27-29], which suggest the use of peripheral vision, were observed. Additionally, the high proportion of non-experts whose ranking scores improved indicates that rather than focusing solely on the specified areas, the use of peripheral vision may have influenced evaluation ability, such as ranking accuracy and scoring improvement.

### 4.4 Future Prospects

In this study, we did not address the need to convey the correctness, the criteria for evaluation and scoring, such as the ideal state of proficient *karate kata* or the influence of sequential patterns of eye movement on evaluation. These aspects require further investigation. Experts and proficient practitioners can identify areas of interest both spatially and temporally and accurately shift their gaze to the next area of interest. Preparing by providing prior knowledge of the focal points used by experts and proficient practitioners will be important to enable participants to observe the necessary areas at the right time. Furthermore, experts and proficient practitioners are likely trained to process information quickly and make evaluation judgments. In contrast, non-experts lack sufficient understanding of key areas and find it difficult to make judgments in a short period. Future experiments should address the elimination of the novelty effect caused by initial exposure to the videos and make it easier to convey correct ranking. For example, experiments could involve only *karate* practitioners as participants, ensuring that they are familiar with the evaluation environment and have an established evaluation framework in their minds before starting the experiment.

Additionally, the relationship between instruction and peripheral vision should be further investigated. In this study, it is hypothesized that peripheral vision, which helps maintain a broad field of view, contributed to the identification of the performers' characteristics and influenced evaluation ability. If it is revealed that gaze-based instruction promotes the use of peripheral vision, it could contribute to passing down the visual exploration strategies of experts in other sports, such as baseball and *Kendo,* where the use of peripheral vision has been identified as a characteristic of proficient practitioners. However, it is difficult to accurately measure the degree of peripheral vision use based on eye movement data alone, as eye tracking estimates gaze position based on the direction of the pupil, and peripheral vision does not necessarily focus on a pinpointed area despite the direction of the pupil. Therefore, additional verification will be needed. Furthermore, clarifying the relationship between peripheral vision and evaluation ability could contribute to the training of instructors.

This study did not examine the combined use of multiple instructional methods. Future experiments should investigate the effects of combining gaze-based and annotation-based instructions, which may help select appropriate instructional methods tailored to specific teaching objectives.

## 5. Conclusion

This study examined the spatiotemporal changes in eye movements caused by two instructional methods: gaze-based instruction and annotation-based instruction. Gaze-based instruction was suggested to potentially promote peripheral vision and possibly contribute to improving evaluation ability. On the other hand, annotation-based instruction seemed likely to encourage participants to observe multiple body parts intentionally, but its effect on improving evaluation ability was suggested to be lower than the gaze-based method. Since gaze-based instruction utilizes the eye movements of experts, it is a versatile instructional method that can be used to convey the tacit knowledge and skills of experts, not only in *karate kata* competitions but also across various fields. Therefore, further detailed investigations of the findings obtained in this study could make it possible to efficiently transfer skills across a wide range of fields, including industry, healthcare, and sports.

**Acknowledgments:**

We extend our special thanks to Mr. Shunpei Takeshi for providing technical support, including assistance with the setup and preparation of the experimental equipment. We are also deeply grateful to Drs. Shinya Chiyohara and Kentaro Hiromitsu for their expert advice on experimental design, key considerations, and data analysis methods. Furthermore, we sincerely thank Mr. Takeshi Kunimoto, a *Shihan* at Nippon Karate-do Taishikan, and all others involved for their cooperation as participants in our experiments. This research was supported by Innovative Science and Technology Initiative for Security (Grant Number JPJ004596), ATLA, Japan, JST Moonshot R&D (Grant Number JPMJMS2034), and the Kayamori Foundation of Informational Science Advancement.

Supplementary Materials:

1. Eye Tracking Results

In main manuscript, Section 3.1 pointed out the fixation duration within the gaze group became shorter after instruction compared to before instruction and also noted a decreasing trend in the number of fixation areas. However, this may be due to the threshold defined in this study for fixation duration being inappropriate for the experimental environment, potentially causing undetected short fixations.

The results of a comparative examination, in which fixation was defined as a state where the eye movement speed remained at or below 600 pixels/sec for at least 80 ms, are shown in Tables S.1 to S.3.

Compared to the table shown in Section 3.1, the mean values for fixation duration, fixation counts, and the number of fixation areas increased in all categories. Subsequently, we examined the trends in changes before and after instruction for each group. While it is not possible to definitively determine whether a fixation duration threshold of 80 ms or 160 ms is more appropriate, the trends remained consistent regardless of the definition used, and the conclusions presented in the main manuscript maintained consistency.

Table S.1 Total Fixation Duration [s]

| Expert 4.84 | Gaze | | Annotation | | Control | |
|---|---|---|---|---|---|---|
| | Before | After | Before | After | Before | After |
| Mean±S.D. | 4.81±1.66 | 4.44±1.87 | 5.12±1.63 | 4.89±1.71 | 4.69±1.82 | 5.13±1.52 |
| Cohens' d | -0.21 | | -0.13 | | 0.26 | |

Table S.2 Fixation counts [-]

| | Instantaneous | | Deliberate | | Summention | |
|---|---|---|---|---|---|---|
| Expert | 17 | | 2 | | 19 | |
| | Before | After | Before | After | Before | After |
| Gaze | 15.57±1.72 | 15.71±3.64 | 1.71±1.11 | 1.57±1.27 | 17.29±2.14 | 17.29±2.81 |
| | 0.05 | | -0.12 | | 0.00 | |
| Annotation | 15.00±5.83 | 14.25±3.81 | 2.25±2.25 | 2.88±1.81 | 17.25±4.33 | 17.13±4.09 |
| | -0.15 | | 0.31 | | -0.03 | |
| Control | 14.50±4.79 | 12.40±5.62 | 2.40±11.84 | 3.40±2.46 | 17.00±4.59 | 15.90±4.25 |
| | -0.40 | | 0.46 | | -0.25 | |

Table S.3 Total number of fixation areas and proportion by area
（excluding the space between both arms and legs）

|  | Total number of fixation areas | | Proportion Arms and Legs [%] | | Proportion Face and Torso [%] | | Proportion Outside of body [%] | | Proportion Saccard [%] | |
|---|---|---|---|---|---|---|---|---|---|---|
| Expert | 7 | | 18.42 | | 18.42 | | 26.84 | | 36.32 | |
|  | Before | After | Before | After | Before | After | Before | After | Before | After |
| Gaze | 6.57 ±1.40 | 5.57 ±1.13 | 19.47 ±7.17 | 13.08 ±10.59 | 27.22 ±19.08 | 27.07 ±20.69 | 16.77 ±9.78 | 18.57 ±10.04 | 36.54 ±21.68 | 41.28 ±24.67 |
|  | -0.79 | | -0.71 | | -0.01 | | 0.18 | | 0.20 | |
| Annotation | 6.30 ±1.19 | 6.75 ±2.43 | 14.61 ±5.97 | 17.96 ±11.14 | 34.93 ±18.39 | 26.18 ±19.28 | 17.96 ±11.75 | 20.53 ±12.00 | 32.50 ±21.57 | 35.33 ±22.29 |
|  | 0.20 | | 0.38 | | -0.46 | | 0.22 | | 0.13 | |
| Control | 5.00 ±1.76 | 5.70 ±1.16 | 7.63 ±9.37 | 11.63 ±8.48 | 37.16 ±22.93 | 35.84 ±17.15 | 17.11 ±17.93 | 20.32 ±14.21 | 38.11 ±23.93 | 32.21 ±20.13 |
|  | 0.47 | | 0.45 | | -0.07 | | 0.20 | | -0.27 | |

Table S.4 Proportion by area （including the space between both arms and legs）

|  | Proportion Arms and Legs [%] | | Proportion Face and Torso [%] | | Proportion Outside of body [%] | | Proportion Saccard [%] | |
|---|---|---|---|---|---|---|---|---|
| Expert | 45.26 | | 18.42 | | 17.89 | | 36.32 | |
|  | Before | After | Before | After | Before | After | Before | After |
| Gaze | 45.94 ±19.16 | 38.50 ±16.95 | 17.52 ±8.90 | 20.23 ±12.48 | 8.65 ±7.89 | 6.62 ±5.67 | 36.54 ±21.68 | 41.28 ±24.67 |
|  | -0.41 | | 0.25 | | -0.30 | | 0.30 | |
| Annotation | 43.82 ±14.24 | 44.93 ±20.88 | 23.68 ±13.80 | 19.74 ±13.45 | 9.87 ±8.53 | 10.99 ±10.32 | 32.50 ±21.57 | 35.33 ±22.29 |
|  | 0.06 | | -0.29 | | 0.12 | | 0.12 | |
| Control | 38.53 ±24.90 | 45.21 ±19.52 | 23.37 ±13.87 | 22.58 ±15.46 | 12.63 ±14.66 | 15.53 ±12.34 | 38.11 ±23.93 | 32.21 ±20.13 |
|  | 0.30 | | -0.05 | | 0.21 | | -0.21 | |

Table S.5 Proportion of space between both arms and both legs [%]

|  | Gaze | | Annotation | | Control | |
|---|---|---|---|---|---|---|
|  | Before | After | Before | After | Before | After |
| Mean±S.D. | 8.12±5.29 | 11.95±9.42 | 8.09±6.65 | 9.54±6.00 | 4.47±4.88 | 4.79±3.62 |
| Cohens' d | 0.50 | | 0.23 | | 0.07 | |

## 2. Ranking Results

The results when the video rated 4th by the expert among the four performance videos is excluded are shown in Table S.6. The results when the three evaluators who assigned rankings other than 4th are excluded are shown in Table S.7. Both tables present the scores before and after instructions.

Table S.6 Scores for the ranking determination of each subject
(Exclude the video ranked fourth)

|  | Gaze | | Annotation | | Control | |
|---|---|---|---|---|---|---|
|  | before | after | before | after | before | after |
| S1 | 4 | 2 | 4 | 4 | 4 | 4 |
| S2 | 2 | 0 | 4 | 2 | 4 | 0 |
| S3 | 2 | 4 | 2 | 2 | 2 | 2 |
| S4 | 4 | 2 | 4 | 4 | 2 | 2 |
| S5 | 2 | 0 | 2 | 2 | 4 | 4 |
| S6 | 4 | 2 | 0 | 2 | 4 | 4 |
| S7 | 0 | 4 | 4 | 2 | 2 | 2 |
| S8 |  |  | 4 | 4 | 0 | 4 |
| S9 |  |  |  |  | 4 | 4 |
| S10 |  |  |  |  | 4 | 0 |
| Mean±S.D. | 2.6±1.5 | 2.0±1.6 | 3.0±1.5 | 2.8±1.0 | 3.0±1.4 | 2.6±1.7 |
| Cohen's d | 0.36 | | 0.19 | | 0.26 | |

Table S.7 Scores for the ranking determination of each subject
(Exclude the outlier subjects)

|  | Gaze | | Annotation | | Control | |
|---|---|---|---|---|---|---|
|  | before | after | before | after | before | after |
| S1 | 4 | 2 |  |  | 4 | 4 |
| S2 | 2 | 0 | 4 | 2 |  |  |
| S3 | 2 | 4 | 2 | 2 | 2 | 2 |
| S4 | 4 | 2 | 4 | 4 | 2 | 2 |
| S5 | 2 | 0 | 2 | 2 | 4 | 4 |
| S6 | 4 | 2 | 0 | 2 | 4 | 4 |
| S7 | 0 | 4 | 4 | 2 | 2 | 2 |
| S8 |  |  | 4 | 4 | 0 | 4 |
| S9 |  |  |  |  |  |  |
| S10 |  |  |  |  | 4 | 0 |
| Mean±S.D. | 2.6±1.5 | 2.0±1.6 | 2.9±1.6 | 2.6±1.0 | 2.8±1.5 | 2.8±1.5 |
| Cohen's d | 0.36 | | 0.22 | | 0.00 | |